# Metasurface-mediated anisotropic radiative heat transfer between nanoparticles

Yong Zhang[1,2], Mauro Antezza[3,4], Hong-Liang Yi[1,2,*], and He-Ping Tan[1,2]

[1]*School of Energy Science and Engineering, Harbin Institute of Technology, Harbin 150001, P. R. China*
[2]*Key Laboratory of Aerospace Thermophysics, Ministry of Industry and Information Technology, Harbin 150001, P. R. China*
[3]*Laboratoire Charles Coulomb (L2C) UMR 5221 CNRS-Université de Montpellier, F- 34095 Montpellier, France*
[4]*Institut Universitaire de France, 1 rue Descartes, F-75231 Paris, France*

Metasurfaces, the two-dimensional (2D) counterpart of metamaterials, have recently attracted a great attention due to their amazing properties such as negative refraction, hyperbolic dispersion, manipulation of the evanescent spectrum. In this work, we propose a theory model for the near field radiative heat transfer (NFRHT) between two nanoparticles in the presence of an anisotropic metasurface. Specifically, we set the metasurface as an array of graphene strips (GS) since it is an ideal platform to implement any metasurface topology, ranging from isotropic to hyperbolic propagation. We show that the NFRHT between two nanoparticles can not only be significantly amplified when they are placed in proximity of the GS, but also be regulated over several orders of magnitude. In this configuration, the anisotropic surface plasmon polaritons (SPPs) supported by the GS are excited and provide a new channel for the near-field energy transport. We analyze how the conductance between two nanoparticles depends on the orientation, the structure parameters and the chemical potential of the GS, on the particle-surface or the particle-surface distances by clearly identifying the characteristics of the anisotropic SPPs such as dispersion relations, propagation length and decay length. Our findings provide a powerful way to regulate the energy transport in the particle systems, meanwhile in turn, open up a way to explore the anisotropic optical properties of the metasurface based on the measured heat transfer properties.

## I. INTRODUCTION

When two objects are brought in proximity to each other, the radiative heat transfer (RHT) between them may be significantly enhanced in the near field. This near field enhancement is caused by the tunneling effect of evanescent modes, especially when surface modes such as surface plasmon polaritons (SPPs) or surface phonon polaritons (SPhPs), are excited [1–9]. The huge radiative heat flux in the near field opens the door to various applications like thermophotovoltaics [10], thermal rectification [11], information processing [12]. Since a large number of heat fluxes is of critical importance in these appealing applications, the ability to control such near-field radiative heat transfer (NFRHT) has attracted much attention in nanoscale science during the past years [13–19].

---

[*] Corresponding author. Tel.: +86-451-86412674; E-mail address: yihongliang@hit.edu.cn

Typically, a remarkable theoretical effort in this domain has been devoted to the study of RHT between two or more particles [20–27]. One of the most popular simplifications is the dipole approximation where NFRHT are computed for point-like particles. This assumption considerably simplifies the calculations. Previous works have been focused on the active control of the cooling and heating of nanoparticles, either in vacuum or in proximity of an interface, as well as of the temperature profile within a collection of nanoparticles. Most Recently, Dong *et al*. [28] and Messina *et al*. [29] addressed the role of surface waves in the energy transport through two or a chain of nanoparticles placed in proximity of a planar interface. It has been shown that the presence of a planar substrate supporting a surface resonance enhances the NFRHT by orders of magnitude at large distances. In addition, Asheichyk *et al*. [30] studied the HT between two nanoparticles placed inside a two-plates cavity. It is found that the presence of plates is not additive in the sense that the results for two plates are distinct from the ones for a single plate studied in Refs. [28,29].

Up to now, the planar substrates considered are those supporting isotropic surface plasmon SPPs or SPhPs). In this work, we focus on the NFRHT between nanoparticles in the presence of an anisotropic metasurfaces. Metasurfaces, the two-dimensional (2D) counterpart of metamaterials, have recently attracted a great attention due to their amazing properties such as negative refraction, hyperbolic dispersion, manipulation of the evanescent spectrum, drastic emission enhancement, cloaking, and electromagnetic transparency, to name a few [31–38]. Unlike the three-dimensional (3D) metamaterials 2D metasurface would enable more ambitious applications with increased resolution and simpler excitation, processing, and retrieval of light via near-field techniques [40,41]. The RHT between planary natural anisotropic materials or patterned structures have been recently studied [42–44]. In this work, we include the contribution of the metasurface by imposing the boundary conditions described by the reflected dyadic Green's function. Specifically, we consider graphene strips (GS) in this work. The homogenization of such a metasurface in the subwavelength approximation ($L<<\lambda$) can be done using the effective medium theory based on the electrostatic approach [45]. It is shown that the GS can propagate plasmons along large distances compared to the plasmons' wavelength [46].

We observe that the anisotropic SPPs supported by the GS has a large effect on NFRHT and can increase these quantities by several orders of magnitude compared to isolated objects. The physics behind this effect is studied both in terms of the distributions of Green's function with respect to the wave-vector and the equal-frequency contours of the light dispersion in the metasurface, in order to well identify the role played by the anisotropic surface mode. Moreover, we show that the NFRHT in our configuration exhibits great tunable features by varying the structures or optical parameters of the GS, and it is related to the modification of the anisotropic SPPs excited on the GS.

The paper is structured as follows. In Sec. II, we introduce the geometry of our system, define the Green's function in the presence of an anisotropic metasurface, and give the expression of the heat flux between the two nanoparticles. Section III introduces the optical properties of the GS, and studies the RHT between two nanoparticles placed in proximity of the metasurface. We show that the presence of GS significantly modify the RHT between two nanoparticles. To get insight to the physical origin of the results we discuss in the wave-vector space the Green's function and in the spatial space the energy density. Section IV and Section V are dedicated to the effects of the interdistance, particle-GS distance and the chemical potential. The propagation length and decay length of the anisotropic SPPs are adopted to interpret the results. Finally, in Sec. VI, we give some conclusive remarks and perspectives.

**II. THEORETICAL ASPECTS**

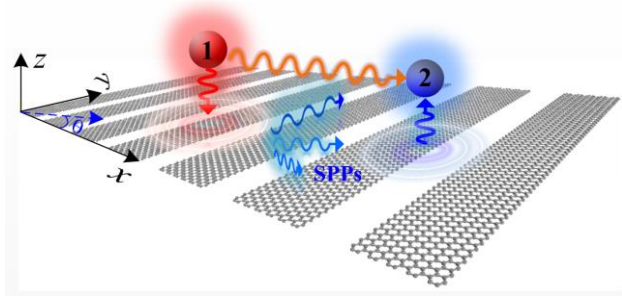

FIG. 1. Radiative heat transfer between two nanoparticles above an array of graphene strips.

To start, let us consider the RHT between two nanoparticles in the presence of a metasurface as shown in Fig. 1. Note that the metasurface shown in Fig. 1 is an array of graphene strips. However, the theory derived below can be applied to any kinds of 2D metasurface. We suppose the nanoparticles are isotropic, linear, nonmagnetic, and the sizes of the nanoparticles are much smaller than the thermal wavelength $\lambda_T = c\hbar/(k_B T)$ so that all individual objects can be modeled to simple radiating electrical dipoles.

The conductance $h$ in the case of two identical nanoparticles can be conveniently expressed in terms of the Green function describing the system as [22]

$$h_{ij} = 3\int_0^{+\infty} \frac{d\omega}{2\pi}\hbar\omega n'(\omega,T) T_{ij}(\omega) \tag{1}$$

where $T_{ij}(\omega) = 4/3 k_0^4 \chi_i \chi_j \mathrm{Tr}\left[\mathcal{G}_{ij}\mathcal{G}_{ij}^*\right]$ is the transmission coefficient (TC) with $k_0 = \omega/c$ and $\chi(\omega) = \mathrm{Im}[\alpha(\omega)] - k_0^3 |\alpha(\omega)|^2/6\pi$ denoting the modified electric frequency-dependent polarizability. In the limit $R \ll \delta$ (with $\delta$ being the skin depth of the given material), $\alpha$ can be written in the well-known Clausius-Mossoti form [47],

$$\alpha(\omega) = 4\pi R^3 \frac{\varepsilon(\omega)-1}{\varepsilon(\omega)+2} \tag{2}$$

with $R$ and $\varepsilon(\omega)$ being the radius and the electric permittivity of the particle, respectively. In this work, we will assume that the two nanoparticles are identical spheres of radius $R = 5$ nm.

$\mathcal{G}$ in Eq. (1) denotes the dyadic Green tensor of the full system, which is written in terms of Green tensor $\mathbb{G}$ as

$$\mathcal{G} = \mathbb{M}^{-1}\mathbb{G} \tag{3}$$

where $\mathbb{M} = \mathbf{I} - k_0^4 \alpha_1 \alpha_2 \mathbb{G}\mathbb{G}^T$ representing the multiple reflections between the two nanoparticles. $n'(\omega,T)$ denotes the derivative with respect to $T$ of the Bose-Einstein distribution $n(\omega,T) = (e^{\hbar\omega/k_B T} - 1)^{-1}$.

As the two nanoparticles are placed on the same side of the metasurface, the Green tensor can be written as,

$$\mathbb{G} = \mathbb{G}_0 + \mathbb{G}_R \tag{4}$$

i.e., separated into a vacuum contribution, and a reflected part which depends on the metasurface reflection matrix and goes to zero in the absence of the metasurface. The vacuum contribution to the Green's function reads,

$$\mathbb{G}_0 = \frac{e^{ik_0 d}}{4\pi k_0^2 d^3} \begin{pmatrix} a & 0 & 0 \\ 0 & b & 0 \\ 0 & 0 & b \end{pmatrix} \tag{5}$$

where $d$ is the distance between the two nanoparticles, $a = 2 - 2ik_0 d$, $b = k_0^2 d^2 + ik_0 d - 1$.

The reflected electric-electric Green's function $\mathbb{G}_{R,EE}$ for the 2D anisotropic surface is expressed as [49,50],

$$\mathbb{G}_{R,EE}(\mathbf{r}_i, \mathbf{r}_j, \omega) = \frac{i}{8\pi^2} \int_{-\infty}^{\infty} dk_x \int_{-\infty}^{\infty} \left( r_{ss}\mathbf{M}_{ss} + r_{ps}\mathbf{M}_{ps} + r_{sp}\mathbf{M}_{sp} + r_{pp}\mathbf{M}_{pp} \right) e^{i\left[k_x|x_i - x_j| + k_y|y_i - y_j|\right]} e^{ik_z|z_i + z_j|} dk_y \tag{6}$$

where $\mathbf{r}_i = x_i \vec{e}_x + y_i \vec{e}_y + z_i \vec{e}_z$,

$$\mathbf{M}_{ss} = \frac{1}{k_z k_\rho^2}\begin{pmatrix} k_y^2 & -k_x k_y & 0 \\ -k_x k_y & k_x^2 & 0 \\ 0 & 0 & 0 \end{pmatrix}, \mathbf{M}_{pp} = \frac{k_z}{k_0^2 k_\rho^2}\begin{pmatrix} -k_x^2 & -k_x k_y & -k_x k_\rho^2/k_z \\ -k_x k_y & -k_y^2 & -k_y k_\rho^2/k_z \\ k_x k_\rho^2/k_z & k_y k_\rho^2/k_z & k_\rho^4/k_z^2 \end{pmatrix},$$

$$\mathbf{M}_{sp} = \frac{1}{k_0 k_\rho^2}\begin{pmatrix} -k_x k_y & -k_y^2 & -k_y k_\rho^2/k_z \\ k_x^2 & k_x k_y & k_x k_\rho^2/k_z \\ 0 & 0 & 0 \end{pmatrix}, \mathbf{M}_{ps} = \frac{1}{k_0 k_\rho^2}\begin{pmatrix} k_x k_y & -k_x^2 & 0 \\ k_y^2 & -k_x k_y & 0 \\ -k_y k_\rho^2/k_z & k_x k_\rho^2/k_z & 0 \end{pmatrix}, \tag{7}$$

and the tensor reflection coefficient $\mathbf{R}$ related to incident '$s$' and '$p$' polarized waves is [45,49]

$$\mathbf{R} = \begin{pmatrix} r_{ss} & r_{sp} \\ r_{ps} & r_{pp} \end{pmatrix}$$

$$= \begin{pmatrix} \dfrac{-\eta_0 \sigma''_{yy}\left(2Z^p + \eta_0 \sigma''_{xx}\right) + \eta_0^2 \sigma''_{xy} \sigma''_{yx}}{\left(2Z^s + \eta_0 \sigma''_{yy}\right)\left(2Z^p + \eta_0 \sigma''_{xx}\right) - \eta_0^2 \sigma''_{xy} \sigma''_{yx}} & \dfrac{-2c^p Z^p \eta_0 \sigma''_{xy}}{\left[\left(2Z^s + \eta_0 \sigma''_{yy}\right)\left(2Z^p + \eta_0 \sigma''_{xx}\right) - \eta_0^2 \sigma''_{xy} \sigma''_{yx}\right]} \\ \dfrac{-2Z^s \eta_0 \sigma''_{yx}}{c^p \left[\left(2Z^s + \eta_0 \sigma''_{yy}\right)\left(2Z^p + \eta_0 \sigma''_{xx}\right) - \eta_0^2 \sigma''_{xy} \sigma''_{yx}\right]} & \dfrac{-\eta_0 \sigma''_{xx}\left(2Z^s + \eta_0 \sigma''_{yy}\right) + \eta_0^2 \sigma''_{xy} \sigma''_{yx}}{\left(2Z^s + \eta_0 \sigma''_{yy}\right)\left(2Z^p + \eta_0 \sigma''_{xx}\right) - \eta_0^2 \sigma''_{xy} \sigma''_{yx}} \end{pmatrix}, \quad (8)$$

being $\eta_0$ the free-space impedance, $Z^s = k_z/k_0$, $Z^p = k_0/k_z$, $c^p = k_z/k_0$ and the identity $k_0^2 = k_x^2 + k_y^2 + k_z^2$ holds. Besides, $\boldsymbol{\sigma}''$ denotes the conductivity tensor in the wave-vector space [45,51],

$$\boldsymbol{\sigma}'' = \begin{pmatrix} \sigma''_{xx} & \sigma''_{xy} \\ \sigma''_{yx} & \sigma''_{yy} \end{pmatrix} = \frac{1}{k_\rho^2} \begin{pmatrix} k_x^2 \sigma'_{xx} + k_y^2 \sigma'_{yy} + k_x k_y \left(\sigma'_{xy} + \sigma'_{yx}\right) & k_x^2 \sigma'_{xy} - k_y^2 \sigma'_{yx} + k_x k_y \left(\sigma'_{yy} - \sigma'_{xx}\right) \\ k_x^2 \sigma'_{yx} - k_y^2 \sigma'_{xy} + k_x k_y \left(\sigma'_{yy} - \sigma'_{xx}\right) & k_x^2 \sigma'_{yy} + k_y^2 \sigma'_{xx} - k_x k_y \left(\sigma'_{xy} + \sigma'_{yx}\right) \end{pmatrix} \quad (9)$$

where $\boldsymbol{\sigma}'$ is the conductivity tensor of the metasurface due to the anticlockwise rotation angle $\theta$ with respect to the spatial coordinates.

$$\boldsymbol{\sigma}' = \mathbf{R}^T \begin{bmatrix} \sigma_{xx} & 0 \\ 0 & \sigma_{yy} \end{bmatrix} \mathbf{R} \quad (10)$$

where $\mathbf{R} = \begin{bmatrix} \cos\theta & -\sin\theta \\ \sin\theta & \cos\theta \end{bmatrix}$.

### III. ANISOTROPIC SPPS MEDIATED HEAT TRANSFER BETWEEN TWO NANOPARTICLES

We now dispose of all the tools needed to calculate the conductance in our system. To gain some insight into the mechanism we want to address, the metasurface in this work is selected as an array of densely packed GS (see Fig. 2(a)) for two reasons: it is an ideal platform to implement any metasurface topology, ranging from isotropic to hyperbolic propagation, and going through the extremely anisotropic $\sigma$-near zero case [41]. Moreover, the chemical potential of graphene and the structure parameters, can affect the optical properties of the surface, offering potential routes toward passive or active control of NFRHT. Here, the effective medium approach (EMA) is adopted to obtain the optical conductivity of the GS. As the strip periodicity W is far less than the plasmons' wavelength $\lambda_{SPPs}$, i.e., W << $\lambda_{SPPs}$, the in-plane effective conductivity tensor $\sigma^{eff}$ of an array of densely-packed GS can be analytically derived using effective medium theory [41] as,

$$\sigma^{eff}_{xx} = \frac{W\sigma\sigma_C}{\sigma_C P + \sigma(W-P)}, \qquad \sigma^{eff}_{yy} = \sigma\frac{P}{W} \quad (11)$$

where P is the ribbon width, $\sigma$ is the graphene conductivity given by [52] and $\sigma_C = -i\omega\varepsilon_0 W/\left(\pi\ln\{\csc[0.5\pi(1-f)]\}\right)$ is an effective conductivity related to the near-field coupling between adjacent strips obtained through an electrostatic approach, in which $f$ is the filling ratio defined as P/W.

As anticipated, we will mainly compare the conductance in vacuum or in the presence of a graphene sheet to the $G$ in the presence of the GS. Concerning the latter, we choose the nanoparticles made of silicon carbide (SiC), a typical polar dielectric materials, the dielectric function of which can be described by the Drude-Lorentz model [53],

$$\varepsilon(\omega) = \varepsilon_\infty \frac{\omega_L^2 - \omega^2 - i\Gamma\omega}{\omega_T^2 - \omega^2 - i\Gamma\omega}, \qquad (12)$$

with high-frequency dielectric constant $\varepsilon_\infty = 6.7$, longitudinal optical frequency $\omega_L = 1.83 \times 10^{14}$ rad/s, transverse optical frequency $\omega_T = 1.49 \times 10^{14}$ rad/s, and damping $\Gamma = 8.97 \times 10^{11}$ rad/s. It is stressed that the expression of the electric polarizability given in Eq. (2) predicts nanoparticle resonance frequency $\omega_r$ corresponds asymptotically to the condition $\varepsilon(\omega) + 2 = 0$, which for SiC gives $\omega_r = 1.755 \times 10^{14}$ rad/s. It is well known that for dielectric nanoparticles, the electric contribution dominates the heat transfer. Thus for SiC nanoparticles, we only consider the electric contribution.

In this section, we consider NFRHT between two nanoparticles. As shown in Figs. 1 and 2(a), we put the GS and nanoparticles on the plane of $xoy$ and the $x$-axis, respectively. Due to the anisotropic structure of the GS, we would examine the rotation angle $\theta$ of the GS on the RHT between nanoparticles, which is defined as the anticlockwise rotation angle with respect to the $x$-axis. Initially, we set the GS parallel to the $y$-axis for $\theta = 0°$. The interparticle and particle-GS distances are denoted by $d$ and $z$, respectively. More importantly, to guarantee the validity of the EMA for our calculations, the particle-GS distance $z$ should be several times greater than the strip periodicity. As shown by Liu *et al*. [14], for W = 20 nm, the EMA predicts the real heat flux well when $z \geq 60$nm.

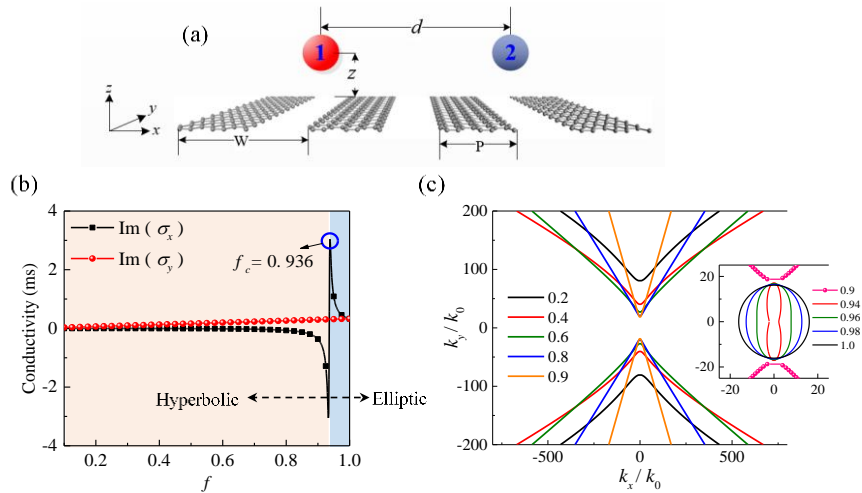

FIG. 2. (a) Schematic of NFRHT between two nanoparticles separated by an interparticle distance of $d$ above an array of GS, where W and P are the strip periodicity and ribbon width, respectively. $z$ is the particle-GS distance for the two nanoparticles. (b) The imaginary part of the optical conductivities along $x$ and $y$ directions with respect to the filling ratio. (c) The dispersion relations of the GS at a rotation angle of $\theta = 0°$ for different filling ratio in wave-vector space at $\omega_r = 1.755 \times 10^{14}$ rad/s.

Let us begin by discussing the optical properties of GS at the $\omega_r$. As a concrete example, the parameter for the GS is chosen as a strip periodicity of W = 20 nm and a chemical potential of $\mu$ = 0.5eV. In Fig. 2(b), we show the imaginary parts of the conductivities Im($\sigma_x$) and Im($\sigma_y$) with respect to the filling ratio $f$. Meanwhile in Fig. 2(c), we plot equal-frequency curves (EFC) of the GS obtained from the dispersion relation [54] $2k_0^2\eta(\sigma_{xx}+\sigma_{yy}) - 2\eta(k_x^2\sigma_{xx}+k_y^2\sigma_{yy}) + k_0 k_z(4+\eta^2\sigma_{xx}\sigma_{yy}) = 0$ being $\eta$ the free-space impedance. We see that at the resonance frequency of SiC, sgn[Im($\sigma_x$)] ≠ sgn[(Im[$\sigma_y$]) is achieved for a large range of $f$ lower than 0.936, indicating a *hyperbolic* topology in the EFC without limitation on the allowed values of $k_x$ and $k_y$, highlighting its unclosed contour that translates into ideally infinitely confined SPPs – i.e., infinite local density of states – propagating towards specific directions within the surface. However we notice a topology transition point at $f$ = 0.936 beyond which sgn[Im($\sigma_x$)] = sgn[(Im[$\sigma_y$]) is achieved providing a *elliptic* anisotropic topology in the EFC. Finally when $f$ = 1, the well known *elliptic* isotropic topology for graphene is realized. As shown in Fig. 2(b), The GS with a larger filling ratio having a larger contrast between Im($\sigma_x$) and Im($\sigma_y$) may favor SPPs propagation towards a specific direction, viz., y-axis in the case. Based on the Fig. 2(c), we can expect that the GS would make great influence on the RHT between two nanoparticles.

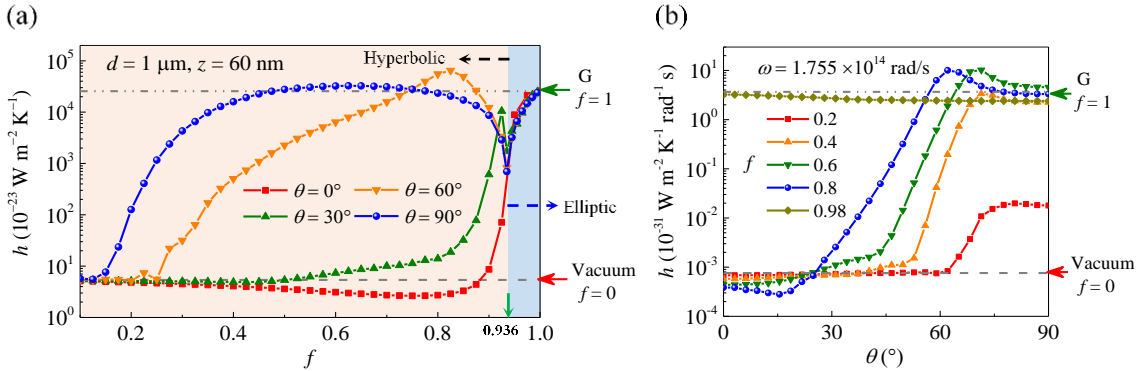

FIG. 3. (a) Total conductance as defined in Eq. (1) between two SiC nanoparticles in the presence of an array of GS at different rotation angle $\theta$ with respect to the filling ratio $f$. (b) Spectral conductance at $\omega_r$ with respect to the rotation angle $\theta$ for different filling ratio $f$. The two gray lines correspond to results in the case of Graphene or vacuum. The particle-GS and interparticle distances are $z$ = 60nm and $d$ = 1μm, respectively.

We first discuss the conductance $h$ at 300K between the two nanoparticles above the GS as a function of the filling ratio $f$ by fixing the particle-GS distance $z$ to the minimum value of 60 nm as shown in Fig. 3(a). The other parameters of GS are the same as those for Fig. 2. We consider a large interparticle distance $d = 1\mu m$. Besides, we also illustrate the impact of the rotation of GS with respect to the reference coordinate system on the results.

We can see in the Fig. 3(a) that the presence of the GS significantly modify the RHT between the two nanoparticles. In most of the cases, the conductance is above that of in the absence of the GS which means that an enhancement of heat transfer is achieved, while is lower than that of in the presence of the graphene sheet. We stress that the enhancement of heat transfer is mainly due to the enhancement of the electromagnetic field on the GS with the excitation of anisotropic SPPs when the two nanoparticles are brought in proximity to the surface. The anisotropic SPPs propagates along the surface, and is coupled to the free-space propagating waves emitted by the nanoparticles, hence providing an additional channel to the energy transportation between the two nanoparticles. As for the NFRHT between two plates made by GS, due to the large wave-vectors of the hyperbolic SPPs the NFRHT are significantly enhanced in comparison with the case of two graphene sheets which support isotropic SPPs with limited wave-vectors [14]. However, this effect is not true in our configuration. When the graphene is patterned into an arrays of GS, a suppression of RHT is observed in most cases as depicted in Fig. 3(a). This can be understood as follows. For the configuration of two plates, the SPPs supporting large wave-vectors is more desirable to generate a giant tunneling of evanescent waves, while the propagation ability along the surface is not important due to the infinite size of the surface. Nevertheless, the propagation characteristic play a great role in our configuration. This means that the propagation length, directionality of the SPPs would make significantly impact on the RHT between two nanoparticles. As the graphene sheet is patterned into strips, the isotropic waves are switched into anisotropic ones. Hence, although the SPPs support large wave-vectors, they propagates towards specific angles, and at the same time is forbidden in other directions as shown in Fig. 2(c). Finally a lower conductance is deduced than that of a graphene sheet as shown in Fig. 3(a).

Now we discuss the results by setting the GS with different filling ratio at different angle with respect to the coordinates. We note that the RHT dominates at the resonant frequency of the SiC particle $\omega_r$ and no other resonance frequencies emerge in our configuration, thus the spectral conductance could predict the trend of the total conductance. Regarding this, we show the $h$ at $\omega_r$ with respect to the rotation angle in Fig. 3(b). We see in Figs. 3(a) and 3(b) that the differences between different angles are orders of magnitude, significantly highlighting the impact of the directionality of the anisotropic SPPs. It is easy to understand that the SPPs prefers to propagation along the graphene strips, viz., $y$ direction for $\theta = 0°$. While SPPs along $x$ direction also exists owing to the resonance coupling of the SPPs between the adjacent strips. Since the two nanoparticles are in-line with $x$-axis, the SPPs along $x$ direction have directly interaction on the RHT between two nanoparticles. In Fig. 2(b), as the filling ratio increases, the dispersion curve get narrower which means the allowed directions become extremely limited, exhibiting even unidirectional SPPs. As a result, a suppression of heat transfer for the case of $\theta = 0°$ gets prominent, and the conductance even lower than the case without GS is observed in Fig. 3(a). Nevertheless, by switch the angle to a larger one, the SPPs propagating towards $x$ direction gets more plentiful, hence an increase of $h$ as shown in Fig. 3(b). A five orders of magnitude enhancement between 90° and 0° at a filling ration of 0.8 is observed. For $\theta > 0°$ in Fig. 3(a), we observe a monotonically increasing behavior as a function of $f$. However, as the filling ratio increases to a large enough one, the ratio between Im($\sigma_x$) and Im($\sigma_y$) gets very large [see in Fig. 2(b)] while the SPPs are still hyperbolic when $f$ is below 0.936. This means that the SPPs is the extremely anisotropic along the GS, hence a decreasing in $h$ is observed. For the GS at a larger angle, the decreasing point appears earlier. Interestingly, the $h$ for 0° increases significantly, switching the suppression effect to an enhancement effect of RHT. This is because that as $f$ increases to a large value, the adjacent strips get very closely. Hence the adjacent SPPs strongly couple with each other through the tunneling effects. We can further observe a turning point at $f = 0.936$ which is in consistence with the turning point as shown in Fig. 3(a) at which the *hyperbolic* topology changes into an *elliptic* anisotropic one in the EFC [see in Fig. 2(c)]. Further increasing the $f$ after this point, *elliptic* anisotropic SPPs gets more like an isotropic one. $h$ thus rises for the GS at any angles, and not surprisingly, the results convergent to those of in the presence of the graphene sheet as the filling ratio goes to 1.0.

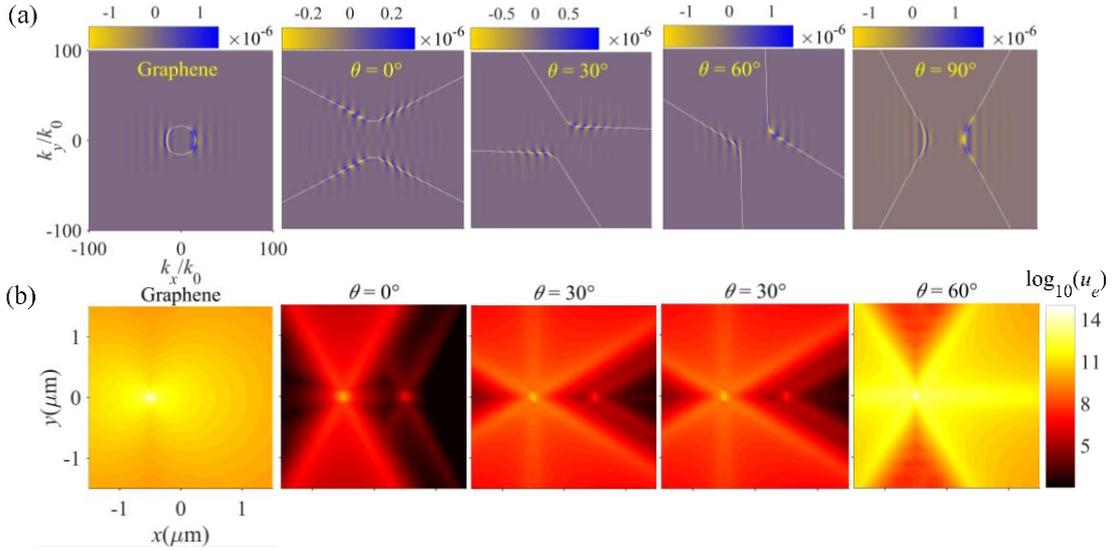

FIG. 4. (a) Wave-vector contours of the real part of the first component of the reflected Green's function $\text{Re}[\mathbf{G}_{R, EE}(1,1)]$ and (b) Spatial contours of the electric field energy density $u_e$ at $z = 30$nm for the graphene sheet and the GS rotated at different angles. In panel (a), the white line corresponds to the equal-frequency curves. For panel (b), the temperatures of the left and right nanoparticles are kept at 300K and 0.5 K, respectively. The interparticle and particle-GS distances are $d = 1$μm and $z = 60$nm, respectively. The frequency is fixed at $\omega_r = 1.755 \times 10^{14}$ rad/s. The filling ratio and chemical potential of the GS are 0.8 and 0.5eV, respectively.

To give an intuitive inspection of the underlying physics, we plot two kinds of contours in the wave-vector space and spatial space, respectively, in Figs. 4(a) and (b). The frequency is chosen as the resonance one, viz., $\omega_r = 1.755 \times 10^{14}$ rad/s. Fig. 4(a) shows the real part of the first component of the reflected Green's function for the graphene and GS at an angle of 0°, 30°, 60° and 90°, We can observe that the isotropic as well as the anisotropic SPPs at different angles are all excited exactly at the dispersion curves, viz., at the resonance wave-vectors. The transitions of the propagation characteristics of the evanescent waves from graphene to GS and from 0° to 90° of GS are also clearly illustrated, confirming our qualitative discussions above. Fig. 4(b) displays the spatial distributions of the radiated electric field energy density

$$\mu_e(\mathbf{r},\omega) = \frac{2\varepsilon_0^2}{\pi\omega}\sum_j \chi_j \Theta(\omega,T_j) \text{Tr}\left[\mathbb{Q}_{rj}\mathbb{Q}_{rj}^*\right] \quad \text{where} \quad \mathbb{Q}_{rj} = \omega^2 \mu_0 \left(G_0^{rj} + G_R^{rj}\right)\mathcal{G} \quad \text{and} \quad \Theta(\omega,T_j) = \hbar\omega n(\omega,T)$$

[28] is the mean energy of the Planck oscillator at the temperature $T$, at the plane of $z = 30$ nm. We see that the presence of the GS significantly modify the energy distributions in the physical space, which exhibit distinctively inhomogeneous while are homogeneous for the case with a graphene sheet. An inspection of the plots at 0° and 90° clearly illustrates that a higher energy density prefers to distribute along the GS. In Fig. 4(a), for a small angle of $\theta = 0°$ or 30°, we see that the EFC and the line $k_y = 0$ do not cross each other as shown, thus few SPPs directly propagate along the x-axis. While the EFC and $k_y = 0$ cross each other for $\theta = 60°$ and 90°, hence a direct propagation channel at x-axis is established. We thus observe a large energy density distributed along x-axis especially for $\theta = 60°$. These physics could give an interpretation for the climbing trend of the blue line in Fig. 3(b).

Interestingly, as depicted in Figs. 3(a) and (b), we see that although the allowed directions for the GS are limited, the conductance of GS can be even larger than that of the graphene and the GS at $\theta = 90°$ when the GS is rotated at a proper angle, i.e., 60° for $f = 0.8$ and 70° for $f = 0.6$ [see in Fig. 3(b)]. This surprising result is a clear indication of the complexities and richness of heat transport in our system. For $f = 0.8$, this enhancement of RHT mainly thanks to the fact the intersection point for $\theta = 60°$ locates at a larger wave-vector than the one for the graphene sheet or the GS at $\theta = 90°$ as shown in Fig. 4(a), hence the propagation of SPPs along x-axis carrying a very large wave-vector is realized. Nevertheless, due to the weak coupling SPPs between adjacent GS for a small filling ratio, this enhancement of RHT can't be realized at any rotation angle, i.e., $f = 0.2$ in Fig. 3(b).

**IV. DISTANCE DEPENDENCE OF THE HEAT TRANSFER**

We now address the question of the dependence of the total conductance regulation on the distances. We stress that we have two distances in our configuration, viz., the interparticle distance $d$ and the particle-surface distance $z$. It is expected that the first distance $d$ is a relevant parameter to highlight the propagation characteristics of the anisotropic SPPs along the surface. While since the surface waves are evanescent waves whose amplitude decreases away from the interface on a wavelength scale, we thus expect that the dependence of the second distance $z$ could be used to elucidate the tunneling effects in our configuration.

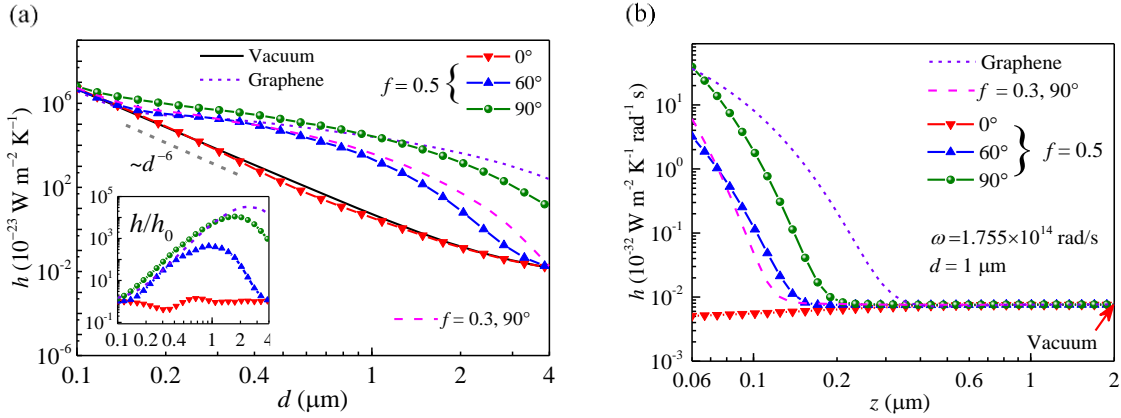

FIG. 5. (a) Total conductance $h$ as defined in Eq. (1) between two SiC nanoparticles at distance $d$ placed at distance $z = 60$ nm from an array of GS. The inset shows the ratio between conductance in the presence and absence of the surface as a function of $d$. (b) Spectral conductance at $\omega_r$ with respect to the particle-surface distance for GS rotated at different angle. The chemical potential of the GS is 0.5eV.

The results of conductance with respect to the first distance $d$ are presented in Fig. 5(a). In the inset, we plot as a function of $d$ the amplification factor $h/h_0$. Eqs. (1) and (5) imply that the small-distance behavior of the conductance in vacuum is $d^{-6}$ [29]. We thus see that the conductance decreases quickly in the absence of the GS. In the presence of GS at a small angle, since the transportation of the evanescent wave along the $x$-direction is blocked, the reflected energy provided by GS is suppressed. Hence, minor differences between the curves with GS at a small angle (30°) and the one in vacuum are observed. While for the GS at a larger angle, although the conductance decreases monotonically, the decay rate is lower than that in the absence of GS. Thus, we see a significantly enhancement of RHT between the two nanoparticles at a large range of interparticle distances as depicted in the inset of Fig. 5(a). The amplification increases monotonically and remarkably reaches a value around $10^4$ for 90° at $d = 1.74$ μm and 400 for 60° at $d = 0.92$ μm, and then turns into a descending trend. We stress that this phenomenon with respect to $d$ is mainly attributed to the propagation length of the SPPs [2], viz., $L=1/\text{Im}(K)$, which can be comparable to one or several wavelengths. $K$ is the resonant parallel wave-vector, determined by the dispersion relations of the surface. Based on the physical meaning of $L$, we can thus expect that in the range of $d < L$, the SPPs excited by the first particle could propagate to the position below the second particle with a big amplitude and then tunneling into it. As for a suspending graphene sheet, $K$ equals to $\sqrt{k_0^2 - (2\varepsilon_0\omega/\sigma)^2}$, implying a propagation length of 1.85μm towards all the directions along the surface at $\omega_r$. We thus observe a decreasing trend in the amplification curve after $d \approx 2$ μm. While we can expect that since the supported SPPs are anisotropic, the propagation length of the SPPs supported by the GS varies with the directions. We find the *L* through calculating the dispersion relations of the GS. The results at $\omega_r$ for the GS rotated at 90° are presented in Fig. 6(a). Not surprisingly, we see that the propagation length validates only at a limited range of directions. Meanwhile, we observe that *L* is large towards the graphene strip and monotonically decreases with respect to the rotation angle. The effective propagate length of GS is thus shorter than that of graphene. Hence we see in Fig. 5(a) the amplification curve decreases at a lower value of *d* than that of graphene. For the GS rotated at an angle less than 90°, due to a shorter propagation length along *x*-axis, the curve also decreases at a shorter distance as shown in Fig. 5(a). Besides, Fig. 6(a) shows that for the GS with a larger filling factor, the *L* curve exhibits a higher value and covers a broader range of propagation directions. We thus see in Fig. 5(a) that the total conductance for *f* = 0.3 decreases faster than that of *f* = 0.5 at the same angle of 90°.

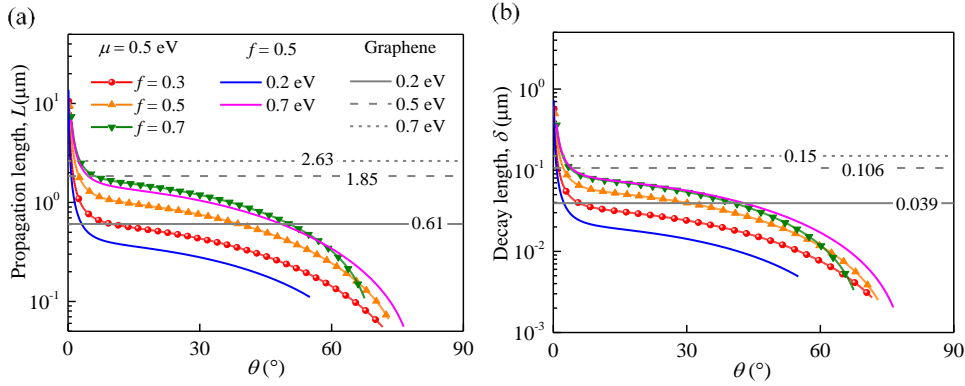

FIG. 6. (a) SPPs propagation length along the surface at $\omega_r$ and (b) SPPs decay length along the *z*-direction at $\omega_r$.

Moreover, we observe in Fig. 5(a) a larger conductance at a small interparticle distance in the presence of GS with *f* = 0.5 than the one in the presence of graphene. This amplification of the flux has also been noticed for *f* = 0.8 as shown in Fig. 3. We stress that this amplification of the flux could be achieved at a large range of interparticle distance once the rotation angle and filling ratio are tuned into proper ones. In order to confirm this point, we show in Fig. 7 the ratio between the spectral conductance at $\omega_r$ of the configuration in the presence of GS and the one in the presence of graphene as a function of *f* with respect to the rotation angle of GS. Note that the white region in Fig. 7 separates the amplification and suppression regions which corresponds to the red and blue colors, respectively. We clearly see that the amplification of flux can be realized at close, middle as well as long distances. The filling ratios of the GS satisfying the amplification are below 0.9, indicating a *hyperbolic* anisotropic SPPs along the GS [see in Fig. 2(a)]. This implies the *hyperbolic* anisotropic SPPs' superiority over the *elliptic* or *isotropic* ones. At a close distance, the higher amplification factor prefers the GS rotated at a larger angle, especially reaches a maximum value at $\theta = 90°$. However, for a larger distance of 500nm or 1μm, the maximum amplification factor increases to a bigger value and is achieved at a smaller rotation angle. Moreover, we see that the green dotted line shows a descending trend versus the $f$, which means a maximum ratio is achieved at a smaller angle for a larger filling ratio. This can be explained that, for a larger filling ratio, the open angle of the dispersion relations is much smaller as shown in Fig. 2(c), hence the intersection between the dispersion curves and $x$-axis could be accomplished at a small angle.

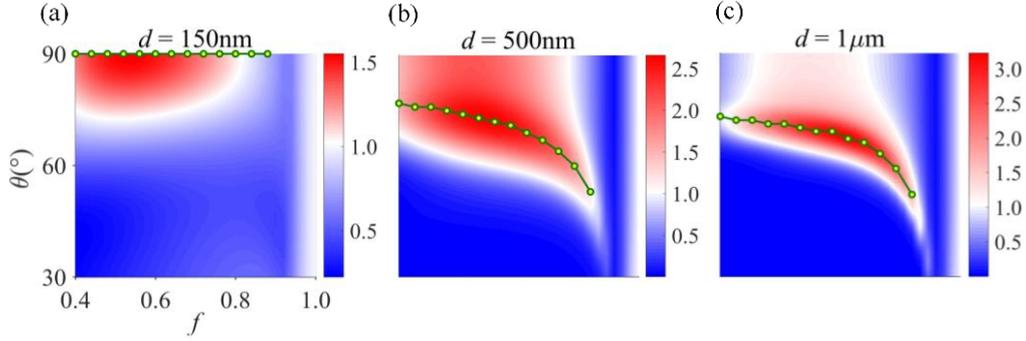

FIG. 7. The ratio between the spectral conductance at $\omega_r$ of the configuration in the presence of GS and the one in the presence of graphene as a function of the filling ratio for the GS with respect to the ration angle of the GS at an interparticle distance of (a) 150nm, (b) 500nm and (c) 1μm. The green symbol lines correspond to the maximum ratio. The chemical potential and the particle-GS distances are 0.5eV and 60nm, respectively.

Now we turn our attention to elucidate the effect of the second distance $z$. Fig. 5(b) shows the spectral conductance at $\omega_r$. We see that, three orders of magnitude enhancement is achieved in the near-field. While with increasing $z$, in other words, from near-field to the far-field, due to the increasingly smaller amplitude of the tunneling evanescent waves, the heat fluxes converge to the one in the absence of the surface. In other words, the enhancement or suppression of heat transfer provided by the surface are negligible in the far-field. To give an explanation on the curves, we plot in Fig. 6(b) the SPPs decay length $\delta = 1/\text{Im}(k_z)$ at $\omega_r$ [2] in the direction perpendicular to the surface. We see that the decay length curves exhibits the same trends as those of the propagation length in Fig. 6(a). The SPPs along the graphene strip possess the longest decay length. We thus see that the heat flux of the GS rotated at 60° converges to a constant value at a shorter $z$ than that of 90°. Due to the decreasing trend of the decay length of the SPPs for the GS, the effective $\delta$ of GS is lower than that of graphene. A faster decay rate in the presence of the GS than that of the graphene is observed in the heat flux curve.

## V. EFFECT OF THE CHEMICAL POTENTIAL

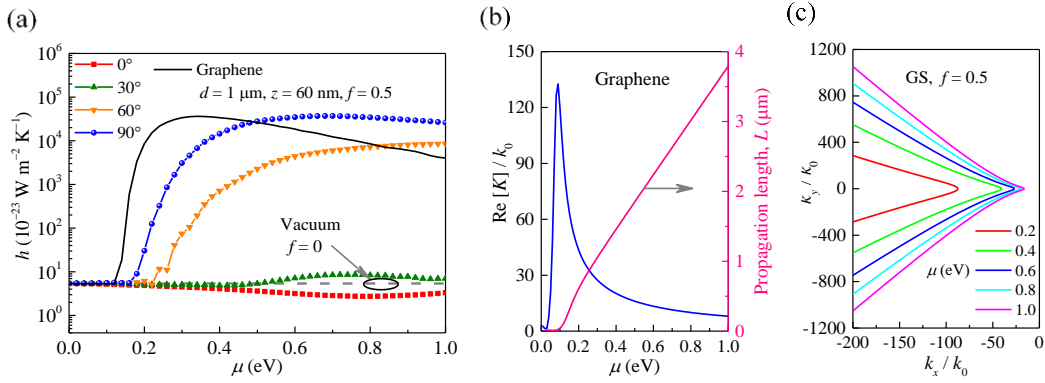

FIG. 8. (a) Total conductance between the two nanoparticles with respect to the chemical potential of the graphene. (b) The lateral wave-vector and propagation length of the graphene sheet. (c) The dispersion relations for the GS. The parameters are kept as $d = 1\,\mu m$, $z = 60$ nm and $f = 0.5$.

It is well known that the chemical potential $\mu$ represents an adjustable parameter allowing us to actively tune the optical properties of graphene. We now examine the influence of the chemical potential on the RHT between nanoparticles. The results for the parameters kept as $d = 1$ μm, $z = 60$ nm and $f = 0.5$ are plotted in Fig. 8(a). To give the explanation of the results, we also present in Fig. 8(b) the dispersion relations $Re[K]/k_0$ and propagation length $L$ for the graphene, and in Fig. 8(c) the dispersion relations for the GS. We see that the impact of the surface on the RHT between nanoparticles is negligible at a very low chemical potential, which results from the near-zero propagation length of the SPPs in graphene though an increasing wave-vector as shown in Fig. 8(b). While we can expect that the propagation length for the GS also increases monotonously with respect to the chemical potential but at a smaller rising rate than that of graphene. The impact of GS thus appears at a larger chemical potential as depicted in Fig. 8(a). We further observe that for the graphene after the impact emerges, with an increase in $\mu$ the curve climbs rapidly and reaches a highest value after which it declines. The climbing trend owes to the increasing propagation length as well as a large enough wave-vector of the SPPs as depicted in Fig. 8(b). However, we can expect that the decline of the wave-vector would reduce the heat flux. With further increasing the chemical potential, this reduction of RHT offsets or even exceeds the increment of RHT contributing from the large propagation length, hence a decline trend in the black curve. As for the GS, although the equal frequency curve moves to the center as $\mu$ increases, it opens and extends to infinite as depicted in Fig. 8(c). As a result, for the GS, the reduction rate of RHT resulting from the decreasing in wave-vector is lower than that of the graphene. We thus see in Fig. 8(a) that the curve for the GS rotated at 90° is above that of the graphene at a large chemical potential. More specifically at 1.0 eV in Fig. 8(a), the conductance for the GS is 6-folds of the one for the graphene. Based on this, we can conclude that for a large chemical potential, the GS is more preferable to amplify the conductance between two nanoparticles to a larger value than the un-patterned graphene. As for the GS rotated at 0°, since the propagation of the SPPs along $x$-axis is forbidden, the increasing propagation length makes no contribution to the RHT. Thus due to the decreasing in wave-vector, the suppressed heat flux gets more dominated at a larger chemical potential.

**VI. CONCLUSIONS**

We have studied the modification of RHT between two dielectric (SiC) nanoparticles when placed in proximity of an array of graphene strips supporting anisotropic SPPs. The SPPs topology of the GS we have used could be tuned from an isotropic one to a hyperbolic one. We have shown that in our configuration, due to the exciting and propagation of the anisotropic surface waves, the ambient-temperature conductance between the nanoparticles can not only be indeed amplified, but also be regulated over several orders of magnitude. We have analyzed the dependence of conductance on the filling factor and the rotation angle of the GS. The underlying physics are interpreted qualitatively through analyzing the distribution of the reflected Green's function in the wave-vector space as well as the spatial electric density distributions, clearly highlighting the role played by the anisotropic surface mode. Moreover, we have studied the dependence of the conductance regulation on the two distances, the interparticle distance $d$ and the particle-surface distance $z$, respectively. We have shown that the enhancement of RHT is significant at a lateral distance comparable to the propagation length of the SPPs. Interestingly, we have also observed that if the rotation angle and filling ratio are tuned into proper ones, the conductance in the presence GS could be larger than the one in the presence of the graphene sheet at close, middle as well as long interparticle distances. This surprising result is a clear indication of the complexities and richness of heat transport in our system. In addition, we have shown that the amplify effect is lost at different large vertical distances for the GS rotated at different angle, as expected since surface waves are confined in the vicinity of the surface. We have adopted the decay length of the SPPs to explain quantitatively these angle dependences. Furthermore, we have shown that the chemical potential of can dramatically modify and allow one to tailor the RHT. We have found that thanks to the hyperbolic topology of the SPPs supported on the GS, at a large chemical potential, the GS is more preferable to amplify the RHT between two nanoparticles to a larger value than the un-patterned graphene.

Our work represents a first step in the study of the modification of energy exchanges mediated by an anisotropic surface and is expected to provide a more powerful way to regulate the energy transport in the particle systems than that by an isotropic surface, meanwhile in turn, opens up a way to explore the anisotropic optical properties of the metasurface based on the measured heat transfer properties. The present study could be extended to examine the radiative heat transport of a chain of nanoparticles or nanoparticle clusters mediated by the anisotropic surface. Meanwhile, the same study could be performed for the nanoparticles placed at each side of a planarly anisotropic slab where the transmission model is needed.

**ACKNOWLEDGMENT**

We thank J. S. Gomez-Diaz and J. Dong for discussions. This work was supported by the National Natural Science Foundation of China (Grant No. 51706053), as well as the Fundamental Research Funds for the Central Universities (Grant No. HIT. NSRIF. 201842), and by the China Postdoctoral Science Foundation (Grant No. 2017M610208).